# THE EFFECTS OF PREDICTION MARKET DESIGN AND PRICE ELASTICITY ON TRADING PERFORMANCE OF USERS: AN EXPERIMENTAL ANALYSIS

Ivo Blohm[1], Christoph Riedl[2], Johann Füller[3], Orhan Köroglu[1], Jan MarcoLeimeister[4], Helmut Krcmar[1]

[1]Technische Universität München
Chair for Information Systems (I17)
Boltzmannstr. 3
85748 Garching b. München, Germany
e-mail: {ivo.blohm, koeroglu, krcmar}@in.tum.de

[2]Harvard University
Institute for Quantitative Social Science
1737 Cambridge St.
Cambridge, MA, 02138, USA
e-mail: criedl@iq.harvard.edu

[3]Universität Innsbruck
School of Management,
Universitätsstraße15
6020 Innsbruck, Austria
e-mail: johann.fueller@uibk.ac.at

[4]Universität Kassel
Chair for Information Systems
Nora-Platiel-Straße 4
34127 Kassel, Germany
e-mail: leimeister@uni-kassel.de

**ABSTRACT**

We employ a 2x3 factorial experiment to study two central factors in the design of prediction markets (PMs) for idea evaluation: the overall design of the PM, and the elasticity of market prices set by a market maker. The results show that 'multi-market designs' on which each contract is traded on a separate PM lead to significantly higher trading performance than 'single-markets' that handle all contracts one on PM. Price elasticity has no direct effect on trading performance, but a significant interaction effect with market design implies that the performance difference between the market designs is highest in settings of moderate price elasticity. We contribute to the emerging research stream of PM design through an unprecedented experiment which compares current market designs.

## INTRODUCTION

The rise of Web 2.0 applications and online communities has empowered firms to tap into the creative potential and knowledge of millions of users. Concepts such as crowdsourcing (Howe, 2008), and open innovation (Chesbrough, 2003) ask for the active engagement of customers, employees, and suppliers in the innovation process. Open innovation (OI) communities have become quite popular. Users from all over the world engage in such platforms and generate numerous ideas, comments, and evaluations. While such online platforms ensure a large variety of ideas and solutions, they bear the problem of evaluating and selecting the best ideas. Various community-based, expert-based and jury-based evaluation and selection methods exist and are applied. However, the selection of the best ideas and prediction of success is still difficult and effortful. Developing precise forecasting tools is of crucial importance since inaccurate evaluation mechanisms imply the risk of selecting the wrong ideas. Lately, prediction markets (PMs) have been introduced as promising tool for collective evaluation tasks. While researchers have demonstrated the appropriateness of PMs during the innovation process, e.g. for the evaluation of new product ideas (Bothos et al., 2009, LaComb et al., 2007, Soukhoroukova et al., 2012), new product concepts (Dahan et al., 2010) and early stage technologies (Chen et al., 2009-10), uncertainty remains about how their design affects their predictive performance.

The predictive performance of PMs depends on certain circumstances such as access to accurate information and to independent knowledge sources which may not hold true for PMs for idea evaluation as they have to deal with high uncertainty and little available information. In this domain, little is known about how the design of the applied PM affects its predictive performance, how it affects traders' skills to cope with these conditions, and how the ideal PM configuration for idea evaluation may look like. In current research, different market designs are applied, i.e. 'single-market designs' which handle all idea contracts on a single market (e.g. Soukhoroukova et al., 2012, Gaspoz/Pigneur, 2008), or 'multi-markets' which set up a separate market for each idea (Bothos et al., 2009) but comparisons of relative performance lack. Additionally, current market makers such as

Logarithmic Market Scoring Rules increase market accuracy (Hanson 2003) and allow to adjust their pricing algorithm such that elasticity of market prices changes. However, it is not yet clear how this market maker is applicable for idea evaluation and how changing price elasticity affects trading performance of users. Thus, the aim of this study is to gain a deeper understanding about the design of PMs for idea evaluation. In detail, we explore: 1) how the two different market designs affect trading performance of PM users and 2) how these effects are moderated by the elasticity of market prices. Answers to these questions shed light on predictive performance of PM for idea evaluation and provide help to configure accurate PMs for idea evaluation. The paper is structured as follows. After a short review of PMs, we present our research design. This is followed by the analysis and a discussion of results, contributions, limitations, and need for future research.

## RELATED WORK

PMs are virtual market places on which users trade contracts that are bound to the occurrence of a future event and whose purpose is to collect, aggregate, and evaluate dispersed information (Wolfers/Zitzewitz, 2004). The theoretical foundation of PMs is the efficient market hypothesis. According to Hayek (1945), market prices are the most efficient instrument to aggregate asymmetrically dispersed information. Thus, market prices in efficient markets can be used for forecasting as they reflect all available information (Fama, 1970). In PMs, traders buy contracts that have a certain payoff (e.g., $100) if a future events occurs. In the case that this event does not occur, contract holders receive no payoff. Hence, the market price reflects the probability that this event occurs, and traders can make profits if they correctly predict the event's occurrence. PMs have successfully been used in the domains of politics, sports, and economics. Researchers also applied the concept to the evaluation of new product ideas (Soukhoroukova et al., 2012, LaComb et al., 2007), new product concepts (Dahan et al., 2010) and early stage technologies (Chen et al., 2009-10).

On PMs, non-binary event spaces, i.e. a magnitude of idea contracts, have been implemented in two different ways. Most researchers set up a single market containing more than two tradable events (Soukhoroukova et al., 2012, Gaspoz/Pigneur, 2008), i.e., all contracts for all ideas are traded on one market. In these markets, traders are able to hold stocks in their portfolio of which they think the underlying event will occur at the market end (we call this 'single-markets'). Contrary, it is also possible to set up a single market for each tradable event or idea ('multi-markets') (Bothos et al., 2009).

In these markets, each idea is represented by two contracts that we call top-contracts ('the idea will be the best idea on the market') and flop-contracts ('the idea will not be the best on the market'). These multiple markets are unified via a common user interface so that they appear as one market to the user. A similar effect can be realized with short-selling functionalities on single-markets (Kamp/Koen, 2009, Wolfers/Zitzewitz, 2004). However, as most members of OI communities will rarely use PM they may lack sufficient knowledge on financial markets in order to apply this complex concept successfully (Blohm et al., 2011).

A major concern of PMs are 'thin markets' in which information aggregation is ineffective due to insufficient traders (Hanson, 2003). Automatic market makers overcome this problem with algorithms that adjust prices based on the transactions of the traders. They give instant feedback to traders, as trades can be performed at any time without having to wait for a second trader as a counterparty (Pennock/Sami, 2008). Thus, market makers add infinite liquidity to PMs. Hanson's (2003) Logarithmic Market Scoring Rules (LMSR) maker is currently the most applied market maker (Jian/Sami, 2012, Slamka et al., 2012). As the LMSR market maker, most market makers apply some kind of mechanism for adjusting its pricing algorithm and its effective liquidity or price elasticity, that can be defined as the degree prices for a given contract change due to a single transaction. In this regard, the LMSR market maker applies an elasticity constant $b$, whose values can be chosen freely.

## RESEARCH MODEL AND HYPOTHESES

Market and contract design are pivotal drivers of accuracy of PMs as it directly affects how dispersed information of multiple traders is aggregated by the market (Wolfers/Zitzewitz, 2004). However, the performance of PMs is not only influenced by how efficient the mechanism aggregates dispersed information, but is also driven by the trading decisions of its users. Appropriately designed PMs may help traders to distinguish more exactly between the single traits of the traded contracts, i.e. a better understanding of idea quality on PMs for idea evaluation, and to convey these judgments into more accurate trading decisions.

We belief that multi-market designs enhance the trading performance of users on PMs for idea evaluation due to two reasons. Firstly, multi-markets could counter the negative effects of 'favorite longshot biases' (Snowberg/Wolfers, 2010). This bias occurs in betting and prediction markets as individuals tend to overbet longshots and underbet favorites. This effect has found to be robust and is grounded in cognitive errors in human information processing. Individuals cannot merely distinguish between small and tiny probabilities. As a result, both are priced similarly, therefore overpaying the smaller one. Further, people distrust very high probabilities leading to relative underevaluations (Snowberg/Wolfers, 2010). The LMSR market maker considers contract prices as probabilities of occurrence. Thus, market prices of all traded contracts equal 1. In OI communities, PMs have to cope with a big amount of ideas. Single-markets handle all these ideas on a unified market (with one market maker), whereas multi-markets employ a single market for every idea (with an own market maker each). Thus, single-market designs should create a higher amount of 'penny stocks' – idea contracts with very low prices. This bigger amount of low priced contracts on single-markets should decrease the ability of PM users to adequately judge idea quality. Secondly, and more importantly, multi-markets should endow a better decision support for traders. Buying an idea contract on single-markets, users can bet on whether the given idea will be of higher quality than the other ideas on the market. On multi-markets traders can also sort out bad ideas by buying flop-contracts. Whereas top-contracts resemble the contracts traded on single-markets and might be appropriate for betting on high quality ideas, flop-contracts might appeal for ideas that users perceive as bad. Thus, flop-contracts help to reduce pricing errors as prices can actively be driven down (Kamp/Koen, 2009, Wolfers/Zitzewitz, 2004). Additionally, on single-markets every transaction affects market prices of all contracts as a single market maker is used. On multi-markets instead, a single transaction is only affected the prices of the traded contracts counterpart. Thus, users of single-markets have to process more information that changes more dynamically. Cognitive load theory suggests that human information processing capacity is limited and information processing errors occur due to cognitive overload. Thus, these additional information processing demands may lead to situations of cognitive overload hampering the trader's ability to make accurate trading decisions (Blohm et al., 2011). Thus, we assume:

H1: *The market design influences the trading performance of users such that 'multi-markets' lead to higher trading performance than 'single-markets.'*

Automatic market makers adjust prices on a given pricing algorithm. Existing research suggests that the efficiency of such market makers is a function of their price elasticity. If price elasticity is too low, prices of idea contracts hardly change and PMs behave very statically. As a consequence, market performance may drop as information exchange via the market mechanism is limited and not enough information can be collected, aggregated, and distributed via the market mechanism (Pennock/Sami, 2008). Thus, market prices in these markets are not efficient such that they cannot transfer private information from well informed to less informed traders (Hayek, 1945). By contrast, too high price elasticity creates highly volatile markets in which prices change very dynamically (Berg/Proebsting, 2009). High degrees of price elasticity create a more complex trading environment as more extreme situations in which high profits or losses can be generated will occur. Traders will have to process more information during the trading process making PM usage a more complex task. Thus, too high price elasticity might lead to situations of cognitive overload in which decision making quality of PM users decreases (Blohm et al., 2011). Thus, trading performance should follow an inverted u-shape as price elasticity increases. However, we believe that the effects of price elasticity on trading performance are conditional on the superordinate market design. If the market design is ill-fitted to the task at hand, the adjustment to an appropriate price elasticity of the market maker will have only a minor effect on trading performance. In H1 it was hypothesized that single-markets lead to a higher cognitive load than multi-markets as traders have to process more information that also changes more dynamically. Thus, the users of single-markets are generally more endangered by the risk of cognitive overload. If price elasticity increases, the cognitive load of prediction market usage should also rise as markets become more volatile. As cognitive load is an additive concept, users of single-markets should be more susceptible to cognitive overload than users of multi-markets. By contrast, users of multi-markets could initially benefit from increasing price elasticity as this market feedback supports them to make more accurate trading decisions. However, if market elasticity is too high, users of multi-markets will also reach a state of cognitive overload hampering their trading performance. Hence, we assume:

H2: *Price elasticity moderates the effect of market design on trading performance such that the difference in trading performance between 'multi-markets' and 'single-markets' will be higher for moderate elasticity settings and smaller for low and high elasticity settings (inverted u-shape).*

Our research model is depicted in Figure 1.

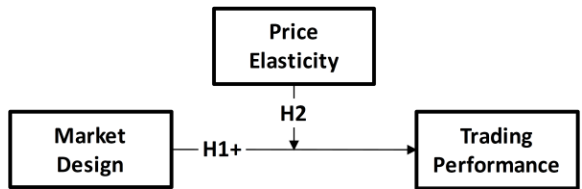
Figure 1: Research Model

## RESEARCH DESIGN

### Experimental Task and Design
In this study six PMs using Hanson's (2003) LMSR market maker are compared. We employ a 2x3 between subject factorial design with random assignment of 323 participants (cf. Table 1). The first factor was market design, where we implemented 'single-market' and 'multi-market' designs. The second factor represented price elasticity, which we varied from low to medium to high. We tested three different price elasticity settings simulating a low (b=877; assuming 80 active traders), a moderate (b=548; assuming 60 active traders), and a high price elasticity on the market (b=219; assuming 40 active traders) in relation to the traders in each market. We used the approach of Berg/Proebsting (2009) to calculate the given degrees of price elasticity.

|  |  | Price Elasticity | | |
|---|---|---|---|---|
|  |  | high | moderate | low |
| Market Design | Single-Market | N = 53 | N = 64 | N = 38 |
|  | Multi-Market | N = 65 | N = 54 | N = 49 |

Table 1: Research Design

We used a standard portal for OI communities developed by the authors for the web experiment. Features, such as idea submissions, or commenting were disabled. Apart from the trading mechanisms all portals were identical. The portal consisted of a summary page containing the ideas to be traded, a portfolio page, and a FAQ explaining the experimental task as well as the PM's way of functioning. The portfolio page contained financial information, such as transaction prices, liquid funds, and a graph representing a trader's overall portfolio value. The participants used their own computers. Before starting the experiment, we confirmed whether all common web browsers displayed the PM correctly. As a web experiment closely reflects the actual usage scenario of OI communities, high external validity of our results can be assumed. Participants could trade the ideas in their natural environment and could allocate as much time to the task as desired. The internal validity of our results was enhanced by analyzing the log files of the PMs. By doing so, inappropriate user behavior, such as a random trading, could be identified. The forecasting goal was set to identify the best five ideas. Intensive pretesting revealed that the subjects perceived the task of identifying the best five ideas as considerably easier than identifying the best idea. On all markets participants received a capital of 5,000 virtual currency units. Participants received a payoff of 100 virtual currency units for each idea contract in their portfolio that were correctly classified and 0 for incorrect classifications.

### Procedure
Based on the random assignment, participants were invited via a personalized email that included a link with the respective system URL and an exhaustive description of the experimental task. Additionally, we provided all participants with a unique activation code that was necessary upon registration on the PM in order to prevent cross-contamination effects and manipulations through the creation of multiple user accounts. The trading period lasted three weeks in November 2010. After the experiment the participants completed a questionnaire.

### Participants
Users of OI communities are predominantly male, young, and well educated (Franke/Shah, 2003). Our sample consisted of undergraduate and graduate students from two information systems courses related to SAP, as well as research assistants from the same field at a large German university. 405 participants took part in the experiment and 323 were included in the analysis. Subjects that did not complete the survey and/or performed two or less trades were removed from the analysis as they did not perform the given trading task adequately. In order to motivate the participants, we offered homework credit points for students and drew two mp3 players for the subjects with the highest trading performance (Slamka et al., 2012). Such rank-order tournament payout schemes were found to enhance accuracy of PMs (Luckner/Weinhardt, 2007). We considered students of the selected SAP courses and information system experts to be appropriate subjects for this study because the experimental task required knowledge of SAP systems. It can also be argued that IS students are suitable participants, as they represent actual users of OI communities. We applied Multivariate Analysis of Variance in order to check random assignment, and found no differences regarding age, gender, and education. There were no differences between students and research assistants. 75.9% of our subjects were male, 5.9% had a master degree, 26.3% a bachelor degree and 60.4% finished high school. Mean age was 22.37 years.

### Idea Sample
The ideas to be evaluated in the experiment comprised of a title and a description. The ideas were taken from a German real-world OI community of the software producer SAP. In this community, SAP users are invited to submit innovative ideas to improve the SAP software. Currently, it consists of 314 users who have submitted 218 ideas varying in length between a half and full A4 page. An independent panel of experts evaluated all ideas. Among all ideas, idea quality is normally distributed (Kolmogorov-Smirnov Z-score: 0.56, p = 0.91). Since conducting an experiment with all ideas implied a substantial workload a stratified sample of 24 ideas was drawn. This sample comprised 8 ideas each with high, medium, and low quality. The sample size was considered sufficient, as 20 to 30 ideas are generally used to measure the variance of creativity ratings of laypersons (Runco/Basadur, 1993).

**Data Sources**

The triangulation of behavioral experiment data, and an expert rating of idea quality helps to gain more robust results overcoming common method bias.

*Behavioral Experiment Data*

The 323 participants performed 12583 transactions. On average, each user performed 38.9 transactions in 93 minutes. We defined trading performance as sum of their disposable cash and their payouts at the market end. We normalized this with the mean number of transactions per market as the absolute number of contracts per market varied on base of the market design.

*Expert Rating*

In practice, companies evaluate innovation ideas with small expert groups (Urban/Hauser, 1993, Girotra et al., 2010). Accordingly, experts are generally used for identifying the most promising ideas in OI communities (Piller/Walcher, 2006). Expert evaluations provide a proxy measure for actual idea quality, which is not observable. Thus, we compared the subjects' transaction with an independent expert evaluation in order to assess their trading performance. Our idea sample was evaluated by a jury using the consensual assessment technique (Amabile, 1996). This technique has been used for evaluating user-generated innovation ideas before (Piller/Walcher, 2006). The jury consisted of 11 referees, who were either professors in information systems, employees of SAP's marketing and R&D department, or the SAP University Competence Centers. Idea quality was measured with four items that are internally used by SAP and reflect the dimensions of novelty, relevance, feasibility, and elaboration as used by Blohm (2011). For evaluation, the idea descriptions were copied into separate evaluation forms which were randomized and contained the scales for idea evaluation as well. The referees were assigned to rate the ideas with the four items on a rating scale from 1 (lowest) to 5 (highest) independently from the other referees with the given forms. We assessed the Intra-Class-Correlation-Coefficients (ICC) of the expert evaluations that should exceed the value of 0.7 (Amabile, 1996). We considered this as met for all items excluding feasibility whose ICC was 0.5 for which ICCs tend to be very low (Amabile, 1996). Based on the mean quality scores of the ideas, we calculated an aggregated quality ranking.

**RESULTS**

**Market level analysis**

In order to test the performance of our markets, we firstly analyzed their accuracy on an aggregated level (cf. Table 2). We checked whether there is a significant concurrence between the markets and the experts, calculating Kendall-Tau rank-order correlations, and Mean Absolute Percentage Errors (MAPE) (Armstrong/Collopy, 1992) using the ranking of ideas according to their final prices at the end of the trading period as well as the ranking of the ideas deriving from the expert evaluation. For multi-markets, we used the prices for top contracts. MAPE is the most widely used measure for evaluating the accuracy of forecasts in time series analysis and offers good validity. We used the placement numbers of the market ranking (forecast ranking) and the expert ranking (actual ranking) for calculating the MAPE (cf. Formula 1). The MAPE thus compares the results of the market outcome with the expert rating. The smaller the MAPE is, the smaller is the market's deviation from the experts.

$$MAPE = \frac{1}{n} \sum_{t=1}^{n} \left| \frac{actual\ ranking\ of\ idea - forecast\ ranking\ of\ idea}{actual\ ranking\ of\ idea} \right|$$

*Formula 1: Mean Absolute Percentage Error*

Generally, the markets tend to correlate stronger among each other than with the expert evaluation. This indicates that they produce quite similar idea rankings. However, the multi-market design with moderate price elasticity (PM5) has the highest correlation (p < 0.05), and the third lowest MAPE, only slightly above the smallest MAPE (9%). PM3 seems to be the most accurate market in terms of MAPE. However, it does not significantly correlate with the expert evaluation so that we consider PM5 as the most accurate market.

|  | Single-Market Design | | | Multi-Market Design | | |
|---|---|---|---|---|---|---|
| Price Elasticity | High (PM1) | Medium (PM2) | Low (PM3) | High (PM4) | Medium (PM5) | Low (PM6) |
| PM1 | -- | 0,29* | 0.49** | 0.47** | 0.49** | 0.27 |
| PM2 |  | -- | 0.29* | 0.38* | 0.37** | 0.05* |

| | | | | 0.44** | 0.37** | 0.21** |
|---|---|---|---|---|---|---|
| PM3 | | | -- | 0.44** | 0.37** | 0.21** |
| PM4 | | | | -- | 0.52** | 0.30* |
| PM5 | | | | | -- | 0.19* |
| Experts | -0.01 | 0.14 | 0.22 | 0.02 | 0.33* | 0.03 |
| MAPE | 1.77 | 1.79 | **1.22** | 1.89 | 1.31 | 1.24 |
| *significant with p < 0.05; **significant with p < 0.01 | | | | | | |

*Table 2: Market Level Analysis*

**Hypothesis Testing**

According to Kamis et al (2008) we applied Partial Least Square (PLS) analysis using SmartPLS 2.0 for testing our research model. We operationalized the experimental conditions as dummy variables. Given the three levels of price elasticity, we created two dummies variables applying the coding scheme of Kamis et al. (2008). Thus, price elasticity dummy 1 (PE Dummy 1) reflects the decrease from a high to a moderate price elasticity setting, and price elasticity dummy 2 the decrease from moderate to low respectively (PE Dummy 2). For testing the moderating effect of price elasticity, we calculated two interaction terms multiplying each of the two price elasticity dummies with the market design dummy. The results are shown in Table 3. We tested H1 in step 1 in which the interaction terms were not included. The coefficient for the market design dummy is positive and significant ($\beta = 0.26$; $p < 0.01$). As the single-market design served as reference group, this implies that multi-market designs lead to higher trading performance of users. Thus, H1 was supported. Additionally, no significant main effect of price elasticity on trading performance was detected. In step 2 we added he interaction terms in our model. Both resulting interaction terms are significant with $p < 0.05$ indicating a significant moderation effect. The coefficient for interaction term 1 is positive ($\beta = 0.23$), whereas interaction term 2 has a negative coefficient ($\beta = -0.18$).

| | Step 1 | Step 2 |
|---|---|---|
| Market Design | 0.262** | 0.173* |
| PE Dummy 1 | -0.080 | 0.032 |
| PE Dummy 2 | 0.024 | -0.093* |
| PE Dummy 1 X Market Design | | 0.227 |
| PE Dummy 2 X Market Design | | -0.178 |
| R² | 0,069 | 0,083 |
| *significant with p < 0.05; **significant with p < 0.01 | | |

*Table 2: Results of PLS analysis*

To probe these results we graphed the estimated means in Figure 2. Whereas trading performance is merely affected by price elasticity on single-markets, trading performance follows an inverted u-shape on multi-markets. Thus the performance difference between the two market designs is highest with a moderate price elasticity. Thus, H2 was supported.

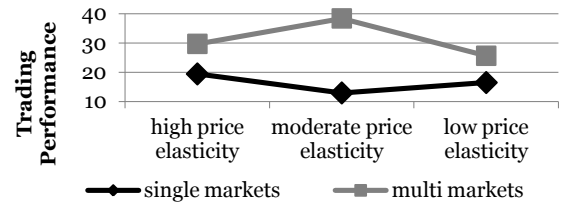

*Figure 2: Estimated means of trading performance*

**DISCUSSION AND CONCLUSION**

Our results suggest that multi-market designs lead to significantly higher market performance than single-market designs and that price elasticity has only a very limited influence on trading performance of PM users on the single-market design. By contrast, the effect of price elasticity on multi-markets is far bigger. On these markets, trading performance follows an inverted u-shape. Decreasing price elasticity from high to moderate levels increases trading performance significantly in the first instance whereas it significantly drops when price elasticity is further diminished to the low elasticity treatment. On average, the multi-market design is 11% more accurate than the single-market design. If moderate price elasticity is considered, this effect rises to 48%. Although the accuracy of prediction markets for idea evaluation can be significantly enhanced using the multi-market design, their evaluation error is quite high. However, the correlation of our market with the expert evaluations is in the range reported by other researchers of 0.43 (LaComb et al., 2007) and of 0.10, 0.39 and 0.47 (Soukhoroukova et al., 2012).

From a theoretical perspective, our work contributes to the growing body of literature on the design on PMs (e.g., Wolfers/Zitzewitz, 2004, Luckner/Weinhardt, 2007), in particular in the domain of idea evaluation. Our work is, to our knowledge, a first-ever experimental analysis of two key factors influencing the performance of PMs for idea evaluation: their fundamental market design (single-markets vs. multi-markets) and price elasticity. Our work shows how these fundamental mechanics interact and how they affect trading performance of PM users. Existing research suggests that PMs work well as long as traders understand the contracts they are supposed to trade (Wolfers/Zitzewitz, 2004). Our research extends this existing line of research and suggests that appropriate market and contract design helps PM users to distinguish more exactly between the different properties of the tradable contracts, and equally important, to express these judgments adequately on the PM. Our research suggests that multi-markets are more apt to meet this conditions than single-markets and that they are most effective in settings of moderate price elasticity. In these conditions, the benefits of multi-markets are amplified as traders get appropriate system feedback stimulating their trading performance. From a practical perspective, our research helps practitioners to set up PMs for idea evaluation and more general for contracts describing vague concepts whose occurrence in future is uncertain. While practitioners are drawn more and more to the concept of PMs, many details necessary to successfully operate markets fur such type of goods are still unavailable. Our work suggests actionable design guidelines that should facilitate application of PMs. In this vein, we suggest to use multi-market designs or design PMs in such manner that traders are provided with sufficient alternatives for each trading decision. Additionally, we suggest a moderate price elasticity. In this regard, we were able to verify the work of Berg/Proebsting (2009) that is useful for calibrating the LMSR market maker.

Some general limitations of controlled experiments apply to our research. While our web-experiment was intended to closely reflect community behavior, general threats to the external validity may result from the use of students. The expert rating might be deficient, although experts generally outperform non-experts (see Ericsson/Lehmann, 1996 for a review). Experts might be more prone to a fixed mind-set rather than a broader community, and thus certain aspects of some ideas might have been overlooked. However, as true idea quality is not directly observable, assessment of idea quality through experts is generally performed for idea selection. The true value of an idea could be considered as the idea's net present value if a value maximizing strategy is pursed (Girotra et al., 2010). However, even after market introduction it can take several years until this value can be determined for a given product (Beardsley/Mansfield, 1978), so that its determination is always associated with high uncertainty as ideas initially submitted in OI communities often merely resemble the final products that have been developed from them. Additionally, the success of products is determined by many different factors beyond idea quality, e.g., the marketing strategy of the focal company. Thus, the accuracy of PMs for idea evaluation resembles the correlation with the community operator's idea selection decisions for a given idea (Kamp/Koen, 2009), that our expert evaluation was intended to approximate. Even when accepting expert judgments as biased, our results retain their validity. Experts are a scarce and valuable resource with limited time. Consequently, using experts for continuously assessing the quality of ideas is expensive. In OI communities a high magnitude of innovation ideas is submitted that cannot be reviewed by experts as it would exceed their resources. Thus, well calibrated PMs can be used to reduce the workload for experts in terms of pre-selection of ideas or even replace the expert panel. Moreover, PMs for idea evaluation suffer from the fact that no real outcome exists, to which payoffs can be tied. This makes additional payout schemes as researched by Slamka et al. (2012) necessary that perform equally well as expert evaluations.

Our study shows how two fundamental mechanics influence the functioning of PMs for idea evaluation. Comparable studies should be replicated in other domains with other types of traded goods. Moreover, our research implies that the used PM designs are perceived cognitively different and these perceptions highly influenced the markets' outcome. Thus, a more indulgent understanding of user cognitions is necessary to design more powerful PMs. In this regard, future research should especially consider the decision process of traders. Understanding this process, markets can be tailored to deliver higher decision support and better market performance. The combination of quantitative and qualitative prediction tools may be a further fruitful avenue for research (Mühlbacher et al., 2011).

## ACKNOWLEDGEMENT

This research received funding through the GENIE project by the German Ministry of Research and Education (BMBF) under contract No. FKZ 01FM07027 and the German Ministry of Economics and Technology (BMWi) under grant code 01MQ07024. The second author also acknowledges supported by the German Research Foundation under grant code RI 2185/1-1.